\pdfoutput=1
\documentclass[11pt,twoside,a4paper,cmspaper,final,collab]{cms-tdr}
\def\svnVersion{13332}\def\cmsCernNo{2010-009}\def\cmsCernDate{2010/05/19}\def\cmsMessage{Published in Physical Review Letters}
\begin{document}\cmsNoteHeader{QCD-10-006}
%
%
%

%
%
\hyphenation{env-iron-men-tal}
\hyphenation{had-ron-i-za-tion}
\hyphenation{cal-or-i-me-ter}
\hyphenation{de-vices}
%
\RCS$HeadURL: svn+ssh://alverson@svn.cern.ch/reps/tdr2/papers/QCD-10-006/trunk/QCD-10-006.tex $
\RCS$Id: QCD-10-006.tex 8191 2010-06-08 11:56:44Z veresg $
%
%
%

\providecommand {\etal}{\mbox{et al.}\xspace} 
\providecommand {\ie}{\mbox{i.e.}\xspace}     
\providecommand {\eg}{\mbox{e.g.}\xspace}     
\providecommand {\etc}{\mbox{etc.}\xspace}     
\providecommand {\vs}{\mbox{\sl vs.}\xspace}      
\providecommand {\mdash}{\ensuremath{\mathrm{-}}} 

\providecommand {\Lone}{Level-1\xspace} 
\providecommand {\Ltwo}{Level-2\xspace}
\providecommand {\Lthree}{Level-3\xspace}

\providecommand{\ACERMC} {\textsc{AcerMC}\xspace}
\providecommand{\ALPGEN} {{\textsc{alpgen}}\xspace}
\providecommand{\CHARYBDIS} {{\textsc{charybdis}}\xspace}
\providecommand{\CMKIN} {\textsc{cmkin}\xspace}
\providecommand{\CMSIM} {{\textsc{cmsim}}\xspace}
\providecommand{\CMSSW} {{\textsc{cmssw}}\xspace}
\providecommand{\COBRA} {{\textsc{cobra}}\xspace}
\providecommand{\COCOA} {{\textsc{cocoa}}\xspace}
\providecommand{\COMPHEP} {\textsc{CompHEP}\xspace}
\providecommand{\EVTGEN} {{\textsc{evtgen}}\xspace}
\providecommand{\FAMOS} {{\textsc{famos}}\xspace}
\providecommand{\GARCON} {\textsc{garcon}\xspace}
\providecommand{\GARFIELD} {{\textsc{garfield}}\xspace}
\providecommand{\GEANE} {{\textsc{geane}}\xspace}
\providecommand{\GEANTfour} {{\textsc{geant4}}\xspace}
\providecommand{\GEANTthree} {{\textsc{geant3}}\xspace}
\providecommand{\GEANT} {{\textsc{geant}}\xspace}
\providecommand{\HDECAY} {\textsc{hdecay}\xspace}
\providecommand{\HERWIG} {{\textsc{herwig}}\xspace}
\providecommand{\HIGLU} {{\textsc{higlu}}\xspace}
\providecommand{\HIJING} {{\textsc{hijing}}\xspace}
\providecommand{\IGUANA} {\textsc{iguana}\xspace}
\providecommand{\ISAJET} {{\textsc{isajet}}\xspace}
\providecommand{\ISAPYTHIA} {{\textsc{isapythia}}\xspace}
\providecommand{\ISASUGRA} {{\textsc{isasugra}}\xspace}
\providecommand{\ISASUSY} {{\textsc{isasusy}}\xspace}
\providecommand{\ISAWIG} {{\textsc{isawig}}\xspace}
\providecommand{\MADGRAPH} {\textsc{MadGraph}\xspace}
\providecommand{\MCATNLO} {\textsc{mc@nlo}\xspace}
\providecommand{\MCFM} {\textsc{mcfm}\xspace}
\providecommand{\MILLEPEDE} {{\textsc{millepede}}\xspace}
\providecommand{\ORCA} {{\textsc{orca}}\xspace}
\providecommand{\OSCAR} {{\textsc{oscar}}\xspace}
\providecommand{\PHOTOS} {\textsc{photos}\xspace}
\providecommand{\PROSPINO} {\textsc{prospino}\xspace}
\providecommand{\PYTHIA} {{\textsc{pythia}}\xspace}
\providecommand{\SHERPA} {{\textsc{sherpa}}\xspace}
\providecommand{\TAUOLA} {\textsc{tauola}\xspace}
\providecommand{\TOPREX} {\textsc{TopReX}\xspace}
\providecommand{\XDAQ} {{\textsc{xdaq}}\xspace}

\providecommand {\DZERO}{D\O\xspace}     


\providecommand{\de}{\ensuremath{^\circ}}
\providecommand{\ten}[1]{\ensuremath{\times \text{10}^\text{#1}}}
\providecommand{\unit}[1]{\ensuremath{\text{\,#1}}\xspace}
\providecommand{\mum}{\ensuremath{\,\mu\text{m}}\xspace}
\providecommand{\micron}{\ensuremath{\,\mu\text{m}}\xspace}
\providecommand{\cm}{\ensuremath{\,\text{cm}}\xspace}
\providecommand{\mm}{\ensuremath{\,\text{mm}}\xspace}
\providecommand{\mus}{\ensuremath{\,\mu\text{s}}\xspace}
\providecommand{\keV}{\ensuremath{\,\text{ke\hspace{-.08em}V}}\xspace}
\providecommand{\MeV}{\ensuremath{\,\text{Me\hspace{-.08em}V}}\xspace}
\providecommand{\GeV}{\ensuremath{\,\text{Ge\hspace{-.08em}V}}\xspace}
\providecommand{\TeV}{\ensuremath{\,\text{Te\hspace{-.08em}V}}\xspace}
\providecommand{\PeV}{\ensuremath{\,\text{Pe\hspace{-.08em}V}}\xspace}
\providecommand{\keVc}{\ensuremath{{\,\text{ke\hspace{-.08em}V\hspace{-0.16em}/\hspace{-0.08em}}c}}\xspace}
\providecommand{\MeVc}{\ensuremath{{\,\text{Me\hspace{-.08em}V\hspace{-0.16em}/\hspace{-0.08em}}c}}\xspace}
\providecommand{\GeVc}{\ensuremath{{\,\text{Ge\hspace{-.08em}V\hspace{-0.16em}/\hspace{-0.08em}}c}}\xspace}
\providecommand{\TeVc}{\ensuremath{{\,\text{Te\hspace{-.08em}V\hspace{-0.16em}/\hspace{-0.08em}}c}}\xspace}
\providecommand{\keVcc}{\ensuremath{{\,\text{ke\hspace{-.08em}V\hspace{-0.16em}/\hspace{-0.08em}}c^\text{2}}}\xspace}
\providecommand{\MeVcc}{\ensuremath{{\,\text{Me\hspace{-.08em}V\hspace{-0.16em}/\hspace{-0.08em}}c^\text{2}}}\xspace}
\providecommand{\GeVcc}{\ensuremath{{\,\text{Ge\hspace{-.08em}V\hspace{-0.16em}/\hspace{-0.08em}}c^\text{2}}}\xspace}
\providecommand{\TeVcc}{\ensuremath{{\,\text{Te\hspace{-.08em}V\hspace{-0.16em}/\hspace{-0.08em}}c^\text{2}}}\xspace}

\providecommand{\pbinv} {\mbox{\ensuremath{\,\text{pb}^\text{$-$1}}}\xspace}
\providecommand{\fbinv} {\mbox{\ensuremath{\,\text{fb}^\text{$-$1}}}\xspace}
\providecommand{\nbinv} {\mbox{\ensuremath{\,\text{nb}^\text{$-$1}}}\xspace}
\providecommand{\percms}{\ensuremath{\,\text{cm}^\text{$-$2}\,\text{s}^\text{$-$1}}\xspace}
\providecommand{\lumi}{\ensuremath{\mathcal{L}}\xspace}
\providecommand{\Lumi}{\ensuremath{\mathcal{L}}\xspace}
%
\providecommand{\LvLow}  {\ensuremath{\mathcal{L}=\text{10}^\text{32}\,\text{cm}^\text{$-$2}\,\text{s}^\text{$-$1}}\xspace}
\providecommand{\LLow}   {\ensuremath{\mathcal{L}=\text{10}^\text{33}\,\text{cm}^\text{$-$2}\,\text{s}^\text{$-$1}}\xspace}
\providecommand{\lowlumi}{\ensuremath{\mathcal{L}=\text{2}\times \text{10}^\text{33}\,\text{cm}^\text{$-$2}\,\text{s}^\text{$-$1}}\xspace}
\providecommand{\LMed}   {\ensuremath{\mathcal{L}=\text{2}\times \text{10}^\text{33}\,\text{cm}^\text{$-$2}\,\text{s}^\text{$-$1}}\xspace}
\providecommand{\LHigh}  {\ensuremath{\mathcal{L}=\text{10}^\text{34}\,\text{cm}^\text{$-$2}\,\text{s}^\text{$-$1}}\xspace}
\providecommand{\hilumi} {\ensuremath{\mathcal{L}=\text{10}^\text{34}\,\text{cm}^\text{$-$2}\,\text{s}^\text{$-$1}}\xspace}


\providecommand{\zp}{\ensuremath{\mathrm{Z}^\prime}\xspace}


\providecommand{\kt}{\ensuremath{k_{\mathrm{T}}}\xspace}
\providecommand{\BC}{\ensuremath{{\mathrm{B}_{\mathrm{c}}}}\xspace}
\providecommand{\bbarc}{\ensuremath{{\overline{\mathrm{b}}\mathrm{c}}}\xspace}
\providecommand{\bbbar}{\ensuremath{{\mathrm{b}\overline{\mathrm{b}}}}\xspace}
\providecommand{\ccbar}{\ensuremath{{\mathrm{c}\overline{\mathrm{c}}}}\xspace}
\providecommand{\JPsi}{\ensuremath{{\mathrm{J}}\hspace{-.08em}/\hspace{-.14em}\psi}\xspace}
\providecommand{\bspsiphi}{\ensuremath{\mathrm{B}_\mathrm{s} \to \JPsi\, \phi}\xspace}
\providecommand{\AFB}{\ensuremath{A_\text{FB}}\xspace}
\providecommand{\EE}{\ensuremath{\mathrm{e}^+\mathrm{e}^-}\xspace}
\providecommand{\MM}{\ensuremath{\mu^+\mu^-}\xspace}
\providecommand{\TT}{\ensuremath{\tau^+\tau^-}\xspace}
\providecommand{\wangle}{\ensuremath{\sin^{2}\theta_{\text{eff}}^\text{lept}(M^2_\mathrm{Z})}\xspace}
\providecommand{\ttbar}{\ensuremath{{\mathrm{t}\overline{\mathrm{t}}}}\xspace}
\providecommand{\stat}{\ensuremath{\,\text{(stat.)}}\xspace}
\providecommand{\syst}{\ensuremath{\,\text{(syst.)}}\xspace}

\providecommand{\HGG}{\ensuremath{\mathrm{H}\to\gamma\gamma}}
\providecommand{\gev}{\GeV}
\providecommand{\GAMJET}{\ensuremath{\gamma + \text{jet}}}
\providecommand{\PPTOJETS}{\ensuremath{\mathrm{pp}\to\text{jets}}}
\providecommand{\PPTOGG}{\ensuremath{\mathrm{pp}\to\gamma\gamma}}
\providecommand{\PPTOGAMJET}{\ensuremath{\mathrm{pp}\to\gamma + \mathrm{jet}}}
\providecommand{\MH}{\ensuremath{M_{\mathrm{H}}}}
\providecommand{\RNINE}{\ensuremath{R_\mathrm{9}}}
\providecommand{\DR}{\ensuremath{\Delta R}}


\providecommand{\PT}{\ensuremath{p_{\mathrm{T}}}\xspace}
\providecommand{\pt}{\ensuremath{p_{\mathrm{T}}}\xspace}
\providecommand{\ET}{\ensuremath{E_{\mathrm{T}}}\xspace}
\providecommand{\HT}{\ensuremath{H_{\mathrm{T}}}\xspace}
\providecommand{\et}{\ensuremath{E_{\mathrm{T}}}\xspace}
\providecommand{\Em}{\ensuremath{E\hspace{-0.6em}/}\xspace}
\providecommand{\Pm}{\ensuremath{p\hspace{-0.5em}/}\xspace}
\providecommand{\PTm}{\ensuremath{{p}_\mathrm{T}\hspace{-1.02em}/}\xspace}
\providecommand{\PTslash}{\ensuremath{{p}_\mathrm{T}\hspace{-1.02em}/}\xspace}
\providecommand{\ETm}{\ensuremath{E_{\mathrm{T}}^{\text{miss}}}\xspace}
\providecommand{\ETslash}{\ensuremath{E_{\mathrm{T}}\hspace{-1.1em}/}\xspace}
\providecommand{\MET}{\ensuremath{E_{\mathrm{T}}^{\text{miss}}}\xspace}
\providecommand{\ETmiss}{\ensuremath{E_{\mathrm{T}}^{\text{miss}}}\xspace}
\providecommand{\VEtmiss}{\ensuremath{{\vec E}_{\mathrm{T}}^{\text{miss}}}\xspace}

\providecommand{\dd}[2]{\ensuremath{\frac{\mathrm{d} #1}{\mathrm{d} #2}}}

%

\providecommand{\ga}{\ensuremath{\gtrsim}}
\providecommand{\la}{\ensuremath{\lesssim}}
\providecommand{\swsq}{\ensuremath{\sin^2\theta_\mathrm{W}}\xspace}
\providecommand{\cwsq}{\ensuremath{\cos^2\theta_\mathrm{W}}\xspace}
\providecommand{\tanb}{\ensuremath{\tan\beta}\xspace}
\providecommand{\tanbsq}{\ensuremath{\tan^{2}\beta}\xspace}
\providecommand{\sidb}{\ensuremath{\sin 2\beta}\xspace}
\providecommand{\alpS}{\ensuremath{\alpha_S}\xspace}
\providecommand{\alpt}{\ensuremath{\tilde{\alpha}}\xspace}

\providecommand{\QL}{\ensuremath{\mathrm{Q}_\mathrm{L}}\xspace}
\providecommand{\sQ}{\ensuremath{\tilde{\mathrm{Q}}}\xspace}
\providecommand{\sQL}{\ensuremath{\tilde{\mathrm{Q}}_\mathrm{L}}\xspace}
\providecommand{\ULC}{\ensuremath{\mathrm{U}_\mathrm{L}^\mathrm{C}}\xspace}
\providecommand{\sUC}{\ensuremath{\tilde{\mathrm{U}}^\mathrm{C}}\xspace}
\providecommand{\sULC}{\ensuremath{\tilde{\mathrm{U}}_\mathrm{L}^\mathrm{C}}\xspace}
\providecommand{\DLC}{\ensuremath{\mathrm{D}_\mathrm{L}^\mathrm{C}}\xspace}
\providecommand{\sDC}{\ensuremath{\tilde{\mathrm{D}}^\mathrm{C}}\xspace}
\providecommand{\sDLC}{\ensuremath{\tilde{\mathrm{D}}_\mathrm{L}^\mathrm{C}}\xspace}
\providecommand{\LL}{\ensuremath{\mathrm{L}_\mathrm{L}}\xspace}
\providecommand{\sL}{\ensuremath{\tilde{\mathrm{L}}}\xspace}
\providecommand{\sLL}{\ensuremath{\tilde{\mathrm{L}}_\mathrm{L}}\xspace}
\providecommand{\ELC}{\ensuremath{\mathrm{E}_\mathrm{L}^\mathrm{C}}\xspace}
\providecommand{\sEC}{\ensuremath{\tilde{\mathrm{E}}^\mathrm{C}}\xspace}
\providecommand{\sELC}{\ensuremath{\tilde{\mathrm{E}}_\mathrm{L}^\mathrm{C}}\xspace}
\providecommand{\sEL}{\ensuremath{\tilde{\mathrm{E}}_\mathrm{L}}\xspace}
\providecommand{\sER}{\ensuremath{\tilde{\mathrm{E}}_\mathrm{R}}\xspace}
\providecommand{\sFer}{\ensuremath{\tilde{\mathrm{f}}}\xspace}
\providecommand{\sQua}{\ensuremath{\tilde{\mathrm{q}}}\xspace}
\providecommand{\sUp}{\ensuremath{\tilde{\mathrm{u}}}\xspace}
\providecommand{\suL}{\ensuremath{\tilde{\mathrm{u}}_\mathrm{L}}\xspace}
\providecommand{\suR}{\ensuremath{\tilde{\mathrm{u}}_\mathrm{R}}\xspace}
\providecommand{\sDw}{\ensuremath{\tilde{\mathrm{d}}}\xspace}
\providecommand{\sdL}{\ensuremath{\tilde{\mathrm{d}}_\mathrm{L}}\xspace}
\providecommand{\sdR}{\ensuremath{\tilde{\mathrm{d}}_\mathrm{R}}\xspace}
\providecommand{\sTop}{\ensuremath{\tilde{\mathrm{t}}}\xspace}
\providecommand{\stL}{\ensuremath{\tilde{\mathrm{t}}_\mathrm{L}}\xspace}
\providecommand{\stR}{\ensuremath{\tilde{\mathrm{t}}_\mathrm{R}}\xspace}
\providecommand{\stone}{\ensuremath{\tilde{\mathrm{t}}_1}\xspace}
\providecommand{\sttwo}{\ensuremath{\tilde{\mathrm{t}}_2}\xspace}
\providecommand{\sBot}{\ensuremath{\tilde{\mathrm{b}}}\xspace}
\providecommand{\sbL}{\ensuremath{\tilde{\mathrm{b}}_\mathrm{L}}\xspace}
\providecommand{\sbR}{\ensuremath{\tilde{\mathrm{b}}_\mathrm{R}}\xspace}
\providecommand{\sbone}{\ensuremath{\tilde{\mathrm{b}}_1}\xspace}
\providecommand{\sbtwo}{\ensuremath{\tilde{\mathrm{b}}_2}\xspace}
\providecommand{\sLep}{\ensuremath{\tilde{\mathrm{l}}}\xspace}
\providecommand{\sLepC}{\ensuremath{\tilde{\mathrm{l}}^\mathrm{C}}\xspace}
\providecommand{\sEl}{\ensuremath{\tilde{\mathrm{e}}}\xspace}
\providecommand{\sElC}{\ensuremath{\tilde{\mathrm{e}}^\mathrm{C}}\xspace}
\providecommand{\seL}{\ensuremath{\tilde{\mathrm{e}}_\mathrm{L}}\xspace}
\providecommand{\seR}{\ensuremath{\tilde{\mathrm{e}}_\mathrm{R}}\xspace}
\providecommand{\snL}{\ensuremath{\tilde{\nu}_L}\xspace}
\providecommand{\sMu}{\ensuremath{\tilde{\mu}}\xspace}
\providecommand{\sNu}{\ensuremath{\tilde{\nu}}\xspace}
\providecommand{\sTau}{\ensuremath{\tilde{\tau}}\xspace}
\providecommand{\Glu}{\ensuremath{\mathrm{g}}\xspace}
\providecommand{\sGlu}{\ensuremath{\tilde{\mathrm{g}}}\xspace}
\providecommand{\Wpm}{\ensuremath{\mathrm{W}^{\pm}}\xspace}
\providecommand{\sWpm}{\ensuremath{\tilde{\mathrm{W}}^{\pm}}\xspace}
\providecommand{\Wz}{\ensuremath{\mathrm{W}^{0}}\xspace}
\providecommand{\sWz}{\ensuremath{\tilde{\mathrm{W}}^{0}}\xspace}
\providecommand{\sWino}{\ensuremath{\tilde{\mathrm{W}}}\xspace}
\providecommand{\Bz}{\ensuremath{\mathrm{B}^{0}}\xspace}
\providecommand{\sBz}{\ensuremath{\tilde{\mathrm{B}}^{0}}\xspace}
\providecommand{\sBino}{\ensuremath{\tilde{\mathrm{B}}}\xspace}
\providecommand{\Zz}{\ensuremath{\mathrm{Z}^{0}}\xspace}
\providecommand{\sZino}{\ensuremath{\tilde{\mathrm{Z}}^{0}}\xspace}
\providecommand{\sGam}{\ensuremath{\tilde{\gamma}}\xspace}
\providecommand{\chiz}{\ensuremath{\tilde{\chi}^{0}}\xspace}
\providecommand{\chip}{\ensuremath{\tilde{\chi}^{+}}\xspace}
\providecommand{\chim}{\ensuremath{\tilde{\chi}^{-}}\xspace}
\providecommand{\chipm}{\ensuremath{\tilde{\chi}^{\pm}}\xspace}
\providecommand{\Hone}{\ensuremath{\mathrm{H}_\mathrm{d}}\xspace}
\providecommand{\sHone}{\ensuremath{\tilde{\mathrm{H}}_\mathrm{d}}\xspace}
\providecommand{\Htwo}{\ensuremath{\mathrm{H}_\mathrm{u}}\xspace}
\providecommand{\sHtwo}{\ensuremath{\tilde{\mathrm{H}}_\mathrm{u}}\xspace}
\providecommand{\sHig}{\ensuremath{\tilde{\mathrm{H}}}\xspace}
\providecommand{\sHa}{\ensuremath{\tilde{\mathrm{H}}_\mathrm{a}}\xspace}
\providecommand{\sHb}{\ensuremath{\tilde{\mathrm{H}}_\mathrm{b}}\xspace}
\providecommand{\sHpm}{\ensuremath{\tilde{\mathrm{H}}^{\pm}}\xspace}
\providecommand{\hz}{\ensuremath{\mathrm{h}^{0}}\xspace}
\providecommand{\Hz}{\ensuremath{\mathrm{H}^{0}}\xspace}
\providecommand{\Az}{\ensuremath{\mathrm{A}^{0}}\xspace}
\providecommand{\Hpm}{\ensuremath{\mathrm{H}^{\pm}}\xspace}
\providecommand{\sGra}{\ensuremath{\tilde{\mathrm{G}}}\xspace}
\providecommand{\mtil}{\ensuremath{\tilde{m}}\xspace}
\providecommand{\rpv}{\ensuremath{\rlap{\kern.2em/}R}\xspace}
\providecommand{\LLE}{\ensuremath{LL\bar{E}}\xspace}
\providecommand{\LQD}{\ensuremath{LQ\bar{D}}\xspace}
\providecommand{\UDD}{\ensuremath{\overline{UDD}}\xspace}
\providecommand{\Lam}{\ensuremath{\lambda}\xspace}
\providecommand{\Lamp}{\ensuremath{\lambda'}\xspace}
\providecommand{\Lampp}{\ensuremath{\lambda''}\xspace}
\providecommand{\spinbd}[2]{\ensuremath{\bar{#1}_{\dot{#2}}}\xspace}

\providecommand{\MD}{\ensuremath{{M_\mathrm{D}}}\xspace}
\providecommand{\Mpl}{\ensuremath{{M_\mathrm{Pl}}}\xspace}
\providecommand{\Rinv} {\ensuremath{{R}^{-1}}\xspace} 
%
%
\newcommand {\roots}    {\ensuremath{\sqrt{s}}}
\newcommand {\dndy}     {\ensuremath{dN/dy}}
\newcommand {\dnchdy}   {\ensuremath{dN_{\mathrm{ch}}/dy}}
\newcommand {\dndeta}   {\ensuremath{dN/d\eta}}
\newcommand {\dnchdeta} {\ensuremath{dN_{\mathrm{ch}}/d\eta}}
\newcommand {\dndpt}    {\ensuremath{dN/d\pt}}
\newcommand {\dnchdpt}  {\ensuremath{dN_{\mathrm{ch}}/d\pt}}
\newcommand {\deta}     {\ensuremath{\Delta\eta}}
\newcommand {\dphi}     {\ensuremath{\Delta\phi}}

\newcommand {\pp}    {\mbox{$pp$}}
\newcommand {\ppbar} {\mbox{p\={p}}}
\newcommand {\pbarp} {\mbox{$p\bar{p}$}}
\newcommand {\PbPb}  {\mbox{PbPb }}

\newcommand{\m}{\ensuremath{\,\text{m}}\xspace}

\newcommand {\naive}    {na\"{\i}ve}
\providecommand{\GEANT} {{Geant}\xspace}
\providecommand{\PHOJET} {\textsc{phojet}\xspace}

\def\d{\mathrm{d}}

\providecommand{\PKzS}{\ensuremath{\mathrm{K^0_S}}}
\providecommand{\Pp}{\ensuremath{\mathrm{p}}}
\providecommand{\Pap}{\ensuremath{\mathrm{\overline{p}}}}
\providecommand{\PgL}{\ensuremath{\Lambda}}
\providecommand{\PagL}{\ensuremath{\overline{\Lambda}}}
\providecommand{\PgS}{\ensuremath{\Sigma}}
\providecommand{\PgSm}{\ensuremath{\Sigma^-}}
\providecommand{\PgSp}{\ensuremath{\Sigma^+}}
\providecommand{\PagSm}{\ensuremath{\overline{\Sigma}^-}}
\providecommand{\PagSp}{\ensuremath{\overline{\Sigma}^+}}
\renewcommand{\GEANTfour} {{Geant4}\xspace}

\cmsNoteHeader{QCD-10-006}
\title{Transverse-momentum and pseudorapidity distributions\\
of charged hadrons in \pp\ collisions at $\sqrt{s}$ = 7 TeV}

\author[neu]{The CMS Collaboration}

\date{\today}

\abstract{Charged-hadron transverse-momentum and pseudorapidity
distributions in proton-proton collisions at $\sqrt{s} = 7$~TeV are
measured with the inner tracking system of the CMS detector at the LHC.
The charged-hadron yield is obtained by counting the number of
reconstructed hits, hit-pairs, and fully reconstructed
charged-particle tracks. The combination of the three methods gives
a charged-particle multiplicity per unit of pseudorapidity
\dnchdeta$|_{|\eta| < 0.5} = 5.78\pm 0.01\stat\pm 0.23\syst$
for non-single-diffractive events, higher than predicted by commonly used
models. The relative increase in charged-particle multiplicity from
$\sqrt{s} = 0.9$ to 7~TeV is ($66.1 \pm 1.0 \stat \pm 4.2\syst$)\%.
The mean transverse momentum is measured to be $0.545\pm 0.005\stat\pm
0.015\syst\GeVc$. The results are compared with similar measurements at
lower energies.}

\hypersetup{
pdfauthor={CMS Collaboration},
pdftitle={Transverse-momentum and pseudorapidity distributions of charged
hadrons in \pp\ collisions at sqrt(s) = 7 TeV},
pdfsubject={CMS},
pdfkeywords={CMS, physics, software, computing}}

\maketitle

{\it Introduction.}
Measurements of particle yields and kinematic distributions are an
essential first step in exploring a new energy regime of particle
collisions. Such studies contribute to our understanding of the physics of
hadron production, including the relative roles of soft and hard
scattering contributions, and help construct a solid foundation for
other investigations. In the complicated environment of LHC \pp\
collisions~\cite{Evans:2008zzb}, firm knowledge of the rates and
distributions of inclusive particle production is needed to
distinguish rare signal events from the much larger backgrounds of soft
hadronic interactions. They will also serve as points of reference for the
measurement of nuclear-medium effects in Pb-Pb collisions in the LHC heavy
ion program.

The bulk of the particles produced in \pp\ collisions arise from soft
interactions, which are modeled only phenomenologically. Experimental
results provide the critical guidance for tuning these widely-used models
and event generators. Soft collisions are commonly classified as elastic
scattering, inelastic single-diffractive (SD) dissociation,
double-diffractive (DD) dissociation, and inelastic non-diffractive (ND)
scattering~\cite{Kittel:2005fu}. (Double-Pomeron exchange is treated as DD
in this paper.) All results presented here refer to inelastic
non-single-diffractive (NSD) interactions, and are based on an event
selection that retains a large fraction of the ND and DD events, while
disfavouring SD events.

The measurements focus on transverse-momentum $\pt$ and pseudorapidity
$\eta$ distributions. The pseudorapidity, commonly used to characterize
the direction of particle emission, is defined as
$\eta=-\ln\tan(\theta/2)$, where $\theta$ is the polar angle of the
direction of the particle with respect to the anti-clockwise beam
direction. The count of primary charged hadrons, $N_{\rm ch}$, is defined
to include decay products of particles with proper lifetimes less than
1~cm. Products of secondary interactions are excluded, and a 
per-cent-level correction is applied for prompt leptons.
The measurements reported here are of \dnchdeta\ and
\dnchdpt\ in the pseudorapidity range $|\eta| < 2.4$ and closely follow
our previous analysis of minimum-bias data at lower centre-of-mass
energies of $\sqrt{s} = 0.9$ and 2.36~TeV as reported in
Ref.~\cite{1stpaper}.

The data for this study are drawn from an integrated luminosity of
1.1\,$\mu{\rm b}^{-1}$ recorded with the Compact Muon Solenoid (CMS)
experiment~\cite{JINST} on March 30th, 2010, during the first hour of the
LHC operation at $\sqrt{s}=7$~TeV. These results are the highest
centre-of-mass energy measurements of the \dnchdeta\ and \dnchdpt\
distributions conducted at a particle collider so far and complement the
other recent measurements of the ALICE experiment at
7~TeV~\cite{alice7tev}.

{\it Experimental methods.}
A detailed description of the CMS experiment can be found in Ref.~\cite{JINST}.
The detectors used for the present analysis are the pixel and
silicon-strip tracker (SST), covering the region $|\eta|< 2.5$
and immersed in a 3.8~T axial magnetic field.
The pixel tracker consists of three barrel layers and two
end-cap disks at each barrel end.
The forward calorimeter (HF), which covers the
region $2.9<|\eta|<5.2$, was also used for event selection.
The detailed Monte Carlo simulation (MC)
of the CMS detector response is based on \GEANTfour\ \cite{GEANT4}.

The event selection and analysis methods in this paper
are identical to those used in Ref.~\cite{1stpaper}, where more details
can be found. The inelastic \pp\ collision rate was about 50~Hz. At these
rates, the fraction of events in the data, where
two or more minimum-bias collisions occurred in the same bunch
crossing, is estimated to be less than 0.3\% and was neglected.
Any hit in the beam scintillator counters
(BSC, $3.23<|\eta|<4.65$) coinciding with colliding proton bunches
was used for triggering the data acquisition. A sample mostly
populated with NSD events
was selected by requiring a primary vertex (PV) to be reconstructed with
the tracker, together with at least one HF
tower in each end with more than 3~GeV total energy.
Beam-halo and other beam-background events were rejected as described in
Ref.~\cite{1stpaper}. The remaining fraction of background events in the
data was found to be less than $2\times10^{-5}$. The numbers of events
satisfying the selection criteria are listed in Table~\ref{tab:eventnum}.

\begin{table}[t!]
  \caption{Numbers of events passing the selection cuts. The selection
criteria are applied in sequence, i.e., each line includes the selection
from the previous ones.}
  \label{tab:eventnum}
  \begin{center}
    \begin{tabular}{lcc}
\hline
\hline
Selection       & Number of events \\
\hline
Colliding bunches + one BSC signal & 68\,512 \\
Reconstructed PV                   & 61\,551 \\
HF coincidence                     & 55\,113 \\
Beam-halo rejection                & 55\,104 \\
Other beam-background rejection    & 55\,100 \\
\hline
\hline
    \end{tabular}
  \end{center}
\end{table}

The event selection efficiency was estimated with
simulated events using the
PYTHIA~\cite{Sjostrand:2006za,Moraes:2007rq} and
PHOJET~\cite{Bopp:1998rc, Engel:1995sb} event
generators. The relative event fractions of SD, DD, and ND processes and
their respective event selection efficiencies are listed in
Table~\ref{tab:difffrac}.
The fraction of diffractive events is predicted by the models
to decrease as a function of collision energy, while the selection
efficiency increases. At $\sqrt{s}=7$~TeV, the fraction of SD (DD) events
in the selected data sample, estimated with PYTHIA and PHOJET, are
6.8\% (5.8\%) and 5.0\% (3.8\%), respectively, somewhat higher than at
$\sqrt{s}=0.9$ and 2.36~TeV~\cite{1stpaper}. With PYTHIA, the 
overall correction for the selection efficiency of NSD processes 
and for the fraction of SD events remaining in the data sample 
lowers the measured charged-particle multiplicity by 6\% compared 
with the uncorrected distribution.

\begin{table}[t!]
  \caption{Fractions of SD, DD, ND, and NSD processes obtained from the
PYTHIA~and PHOJET~event generators before any selection, and the
corresponding selection efficiencies determined from the MC simulation.}
  \label{tab:difffrac}
  \begin{center}
    \begin{tabular}{lcc||cc}
\hline
\hline
        & \multicolumn{2}{c||}{PYTHIA} & \multicolumn{2}{c}{PHOJET} \\
\hline
                & Frac. & Sel. eff. & Frac. & Sel. eff. \\
\hline
 SD             & 19.2\%    & 26.7\%    & 13.8\% & 30.7\% \\
 DD             & 12.9\%    & 33.6\%    &  6.6\% & 48.3\% \\
 ND             & 67.9\%    & 96.4\%    & 79.6\% & 97.1\% \\
 NSD            & 80.8\%    & 86.3\%    & 86.2\% & 93.4\% \\
\hline
\hline
    \end{tabular}
  \end{center}
\end{table}

The \dnchdeta\ distributions were obtained, as in Ref.~\cite{1stpaper},
with three methods, based on counting the following quantities:
(i) reconstructed clusters in the barrel part of the pixel detector;
(ii) pixel tracklets composed of pairs of clusters in different pixel
barrel layers; and (iii) tracks reconstructed in the full tracker volume.
The third method also allows a measurement of the \dnchdpt\
distribution. All three methods rely on the reconstruction of a
PV~\cite{Sikler:2009nx}.
The PV reconstruction efficiency was found to be 98.3\% (98.0\%) in data
(MC), evaluated after all other event selection cuts. In case of multiple
PV candidates, the vertex with the largest track multiplicity was chosen.
The three methods are sensitive to the measurement of particles
down to \pt values of about 30, 50, and 100 \MeVc, respectively.
Only 0.5, 1.5, and 5\% of all charged particles are estimated to be
produced below these \pt values, respectively, and these fractions 
were corrected for.

\begin{figure}[t]
\centerline{\includegraphics[width=0.55\columnwidth]{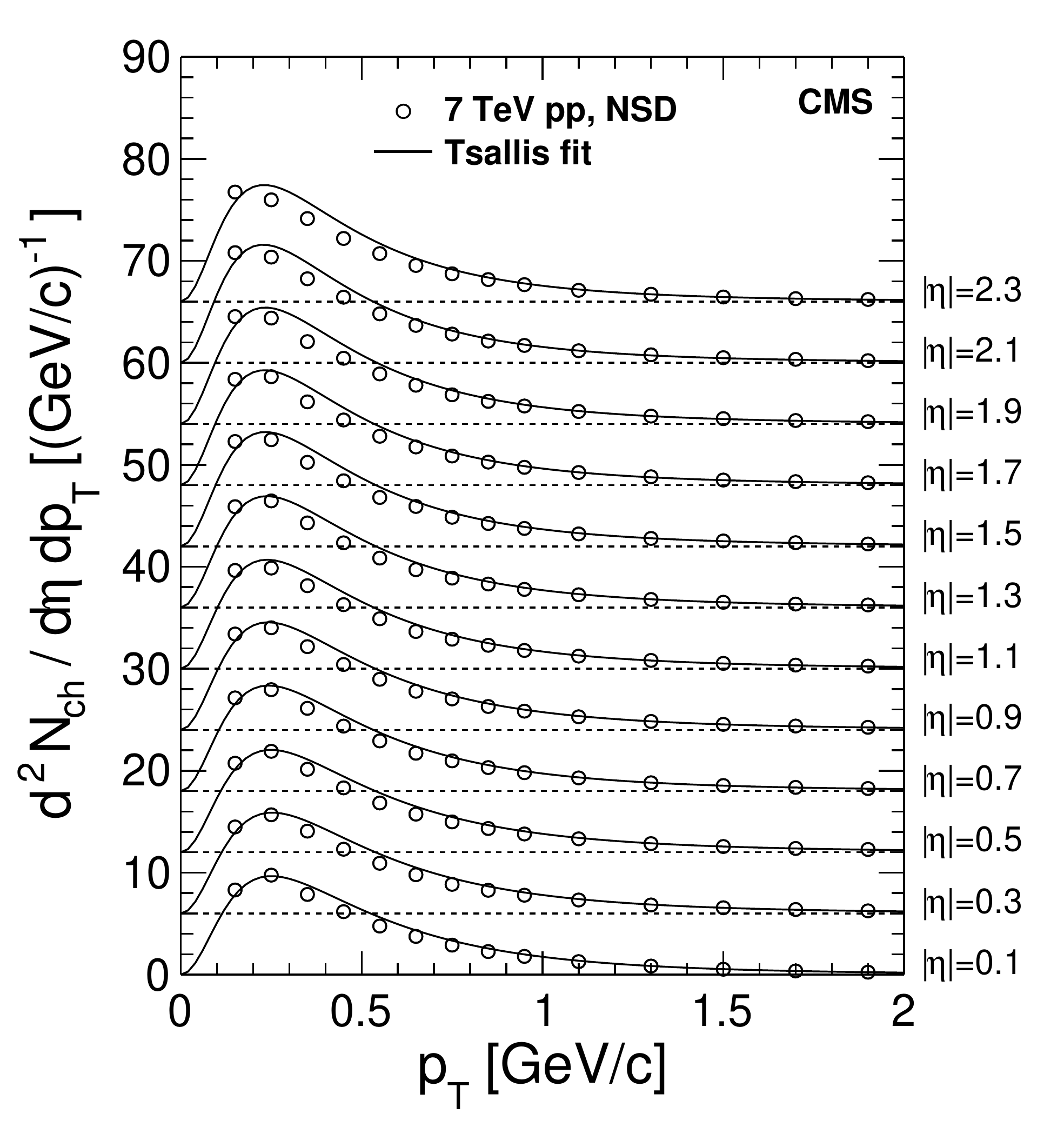}}
\caption{\label{pt_a}
Differential yield of charged hadrons in the range $|\eta| < 2.4$ in
$0.2$-unit-wide bins of $|\eta|$ in NSD events. The solid curves
represent fits of Eq.~\ref{eq:tsallis} to the data.
The measurements with increasing $\eta$ are successively shifted by six
units along the vertical axis.}
\end{figure}

The measurements were corrected for the geometrical acceptance
($\approx$2\%), efficiency ($\approx$5-10\%), fake ($<$1\%) and
duplicate tracks ($<$0.5\%), low-\pt particles curling in the
axial magnetic field ($<$1\%), decay products of long-lived
hadrons ($<$2\%) and photon conversions ($<$1\%) and inelastic
hadronic interactions in the detector material 
($\approx$1-2\%), where the size of the corrections in 
parentheses refer to the tracking method.
The PYTHIA parameter set from
Ref.~\cite{Moraes:2007rq} was chosen to determine the corrections,
because it reproduces the \dnchdeta\ and charged-particle
multiplicity distributions, as well as other control distributions
at 7~TeV, better than other available tuning parameter sets.
Although the corrections do not depend significantly on the model
used, it is indeed important that the simulated data set contains
a sufficient number of high-multiplicity events to determine these
corrections with the desired accuracy.

\begin{figure}[t]
\centerline{\includegraphics[width=0.55\columnwidth]{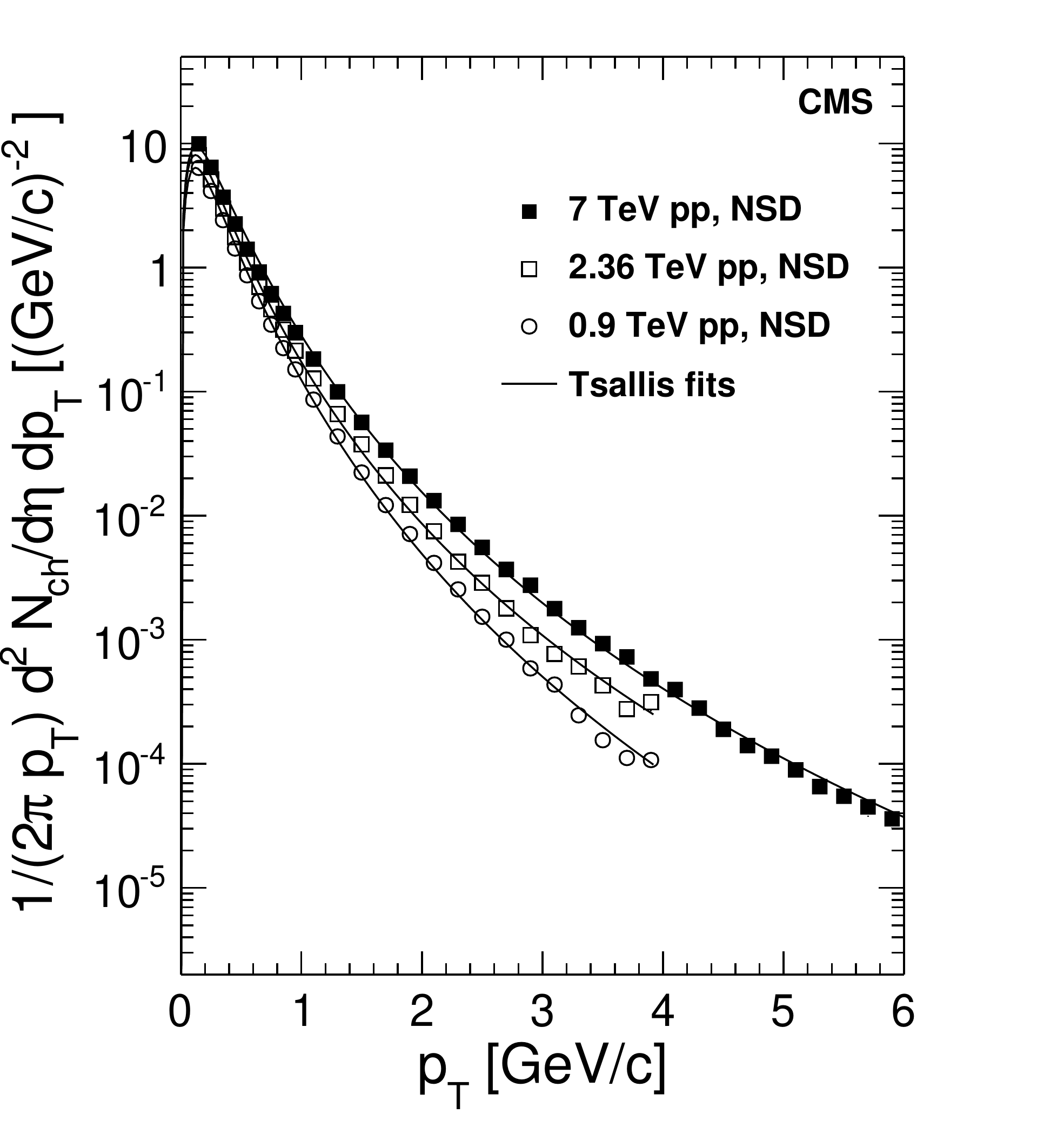}}
\caption{\label{pt_b}
Charged-hadron yield in the range $|\eta| < 2.4$ in NSD events
as a function of \pt; the systematic uncertainties are smaller than the
symbols. The measurements at $\sqrt{s}=0.9$ and 2.36~TeV~\cite{1stpaper}
are also shown. The solid lines represent fits of Eq.~\ref{eq:tsallis}
to the data.}
\end{figure}

{\it Results.}
For the measurement of the \dnchdpt\ distribution, charged-particle tracks
with \pt in excess of 0.1\GeVc were used in 12 different
$|\eta|$ bins, from 0 to 2.4. The average charged-hadron yields in
NSD events are shown in Fig.~\ref{pt_a} as a function of \pt
and $|\eta|$. The Tsallis parametrization
\cite{Tsallis:1987eu,Wilk:2008ue,Biro:2008hz},

\begin{equation}
 E \frac{d^3N_{\mathrm{ch}}}{d p^3} =
\frac{1}{2\pi \pt} \frac{E}{p} \frac{d^2N_{\mathrm{ch}}}{d\eta\,d\pt} =
C \frac{dN_{\mathrm{ch}}}{dy}\left(1 + \frac{E_\mathrm{T}}{nT}\right)^{-n},
 \label{eq:tsallis}
\end{equation}

\noindent
where $y=0.5\ln[(E+p_z)/(E-p_z)]$,
$\ET = \sqrt{m^2 + \pt^2} - m$, and $m$ is the charged pion mass,
was fitted to the data. The \pt\ spectrum of charged hadrons,
$1/(2\pi\pt)d^2N_{\rm ch}/d\eta\,d\pt$, measured in the range
$|\eta| < 2.4$, is shown in Fig.~\ref{pt_b} for data at 0.9, 2.36, and
7~TeV. The high-\pt reach of the data is limited by the increase of
systematic uncertainties with \pt.
The fit to the data (Eq.~\ref{eq:tsallis}) is mainly used for
extrapolations to $\pt =0$, but is not expected to give a good
description of the data in all $\eta$ bins with only two parameters.
The parameter $T$ and the exponent $n$ were found to be
$T = 0.145 \pm 0.005 \syst$~GeV and $n = 6.6 \pm 0.2\syst$. The
average \pt, calculated from a combination of the measured data points and
the low- and high-\pt\ contributions as determined from the fit, is
$\langle \pt \rangle = 0.545 \pm 0.005\stat\pm 0.015\syst\GeVc$.

Experimental uncertainties related to the trigger and event selection
are common to all the analysis methods. The uncertainty related
to the presence of SD (DD) events in the final sample was
estimated to be 1.4\% (1.1\%), based on
consistency checks between data and simulation for diffractive event
candidates. The total event selection uncertainty, which also includes
the selection efficiency of the BSC and HF, was found to be 3.5\%.
Based on studies similar to those presented in Ref.~\cite{1stpaper},
additional 3\% and 2\% uncertainties were assigned to the tracklet and
track reconstruction algorithm efficiencies, respectively.
Corrections at the percent level were applied to the final results to
extrapolate to $\pt = 0$. The uncertainty on these extrapolation
corrections was found to be less than 1\%. All other uncertainties are
identical to those listed in Ref.~\cite{1stpaper}.
The \dnchdeta\ measurements were repeated on a separate data sample
without any magnetic field, for which almost no \pt extrapolation is
needed, and gave results consistent within 1.5\%.
The final systematic uncertainties for the pixel counting, tracklet, and
track methods were found to be 5.7\%, 4.6\%, and 4.3\%,
respectively, and are strongly correlated.

\begin{figure}[t]
\centerline{\includegraphics[width=0.55\columnwidth]{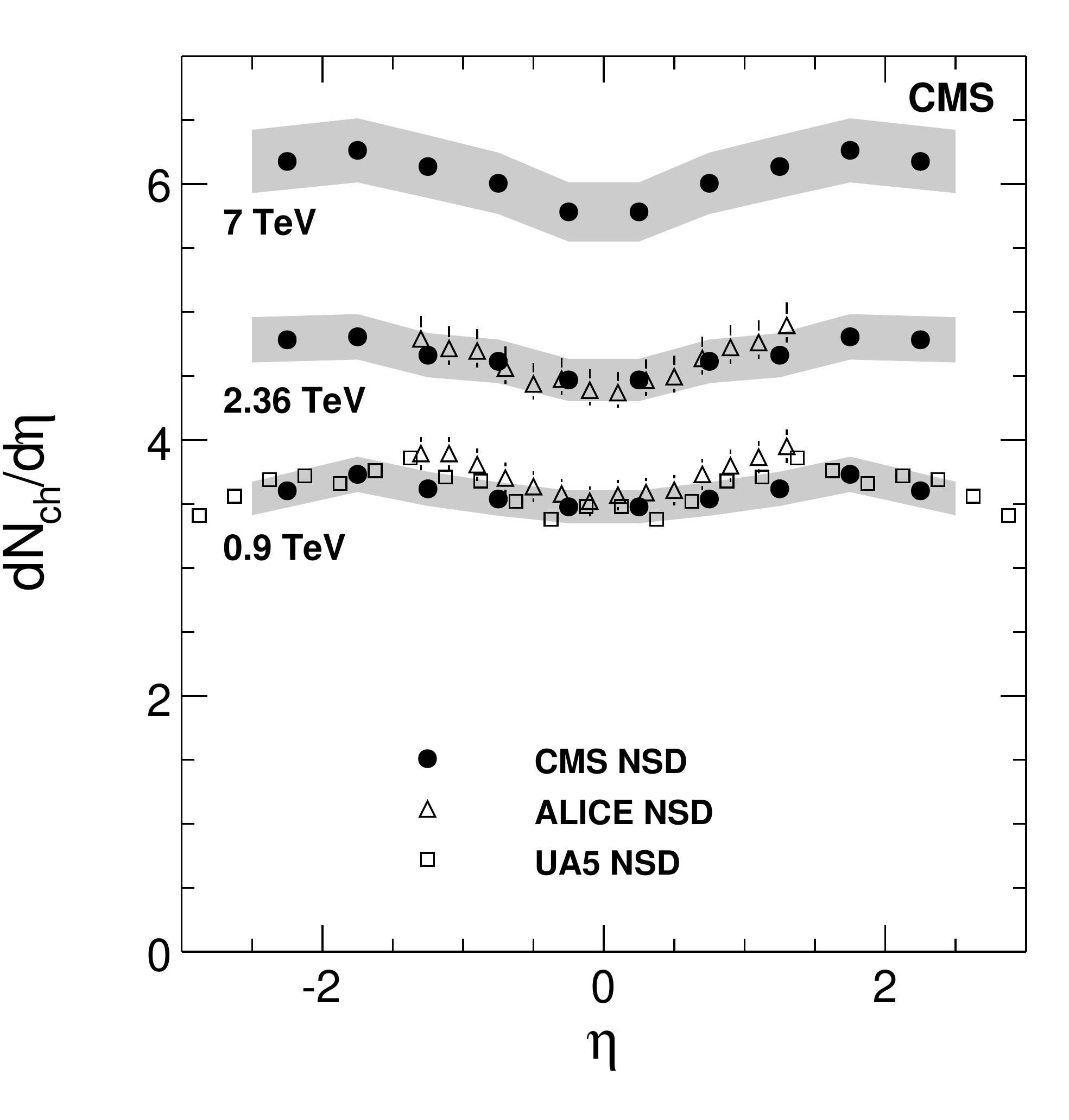}}
\caption{\label{dndeta_b} Distributions of $dN_{ch}/d\eta$, averaged over
the three measurement methods and compared with
data from UA5~\cite{Alner:1986xu} (\pbarp, with statistical errors only)
and ALICE~\cite{alice} (with systematic uncertainties).
The shaded band shows systematic uncertainties of the CMS data. The CMS
and UA5 data are averaged over negative and positive values of $\eta$.}
\end{figure}

For the \dnchdeta\ measurements,
the results for the three individual layers within the cluster-counting
method were found to be consistent within 1.2\% and were combined.
The three layer-pairs in the pixel-tracklet method provided results that
agreed within 0.6\% and were also combined. Finally, the results from the
three different measurement methods, which agree with the combined
result within 1\% to 4\% depending on $\eta$, were averaged. The final
\dnchdeta\ distributions are shown in Fig.~\ref{dndeta_b} for
$\sqrt{s}=0.9$, 2.36, and 7~TeV. The CMS results are compared with
measurements made by other experiments. In the ATLAS Collaboration
analysis~\cite{Aad:2010rd}, events and particles were selected in a
different region of phase space, which makes a direct comparison
difficult. Their results are therefore not included in the figure.

\begin{figure}[t]
\centerline{\includegraphics[width=0.55\columnwidth]{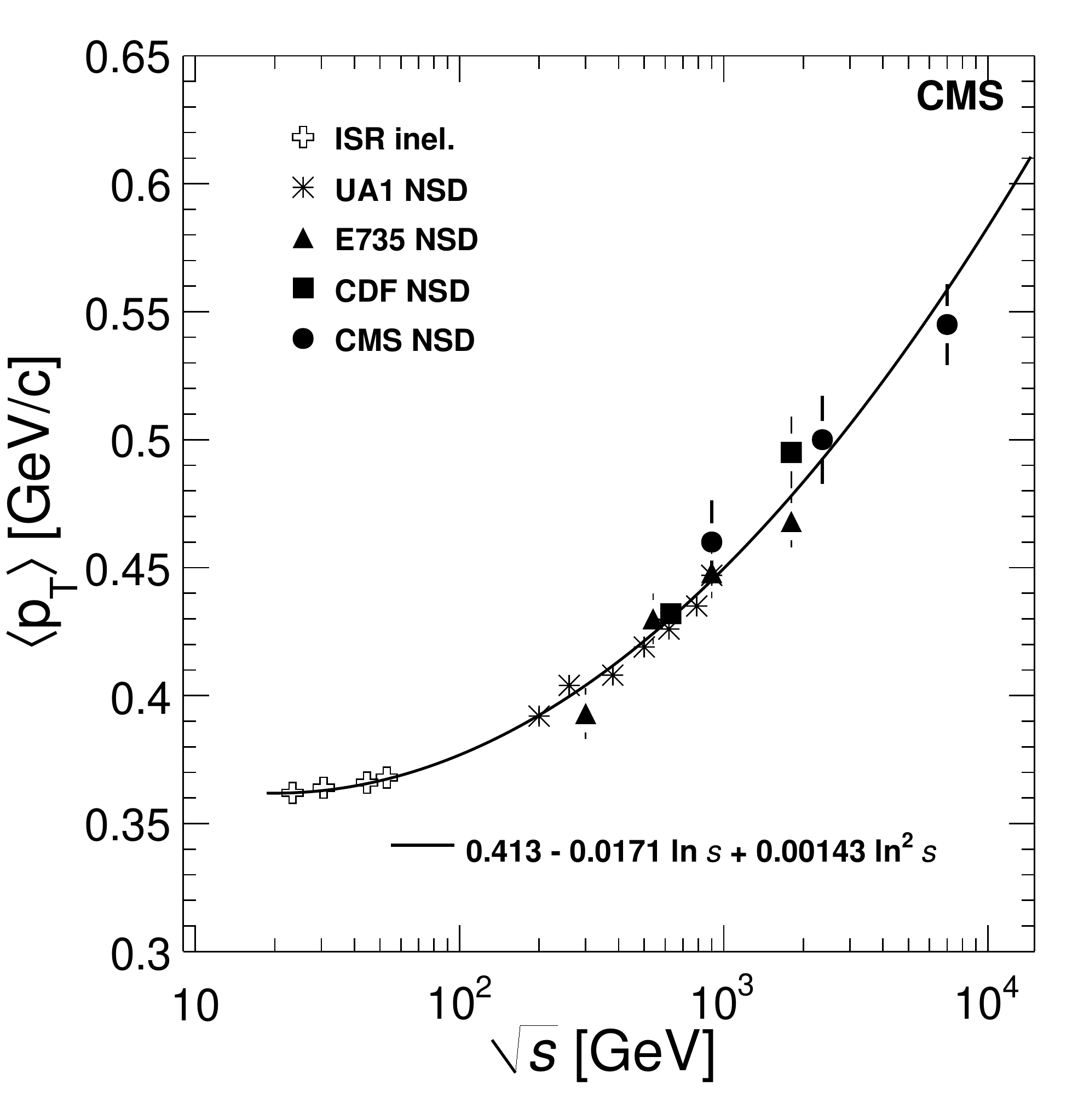}}
\caption{\label{roots_a}
Average \pt of charged hadrons as a function of the centre-of-mass
energy. The CMS measurements are for $|\eta|<2.4$.
Also shown are measurements from the ISR~\cite{Rossi:1974if} (\pp),
E735~\cite{Alexopoulos:1988na} (\pbarp), and CDF~\cite{Abe:1988yu} (\pbarp)
for $|\eta|<0.5$, and from UA1~\cite{Albajar:1989an} (\pbarp) for
$|\eta|<2.5$. The solid line is a fit of the functional form
$\langle \pt \rangle=0.413-0.0171\ln s + 0.00143\ln^2 s$ to the data.
The error bars on the CMS data include the systematic uncertainties.}
\end{figure}

The results can also be compared to earlier experiments as a function of
$\sqrt{s}$. The energy dependence of the average charged-hadron \pt
can be described by a quadratic function of
$\ln{s}$~\cite{Albajar:1989an}. As shown in Fig.~\ref{roots_a}, the
present measurement follows this trend. The choice of the $|\eta|$ 
interval can influence the average \pt value by a few per cent.

\begin{figure}[t!]
\centerline{\includegraphics[width=0.55\columnwidth]{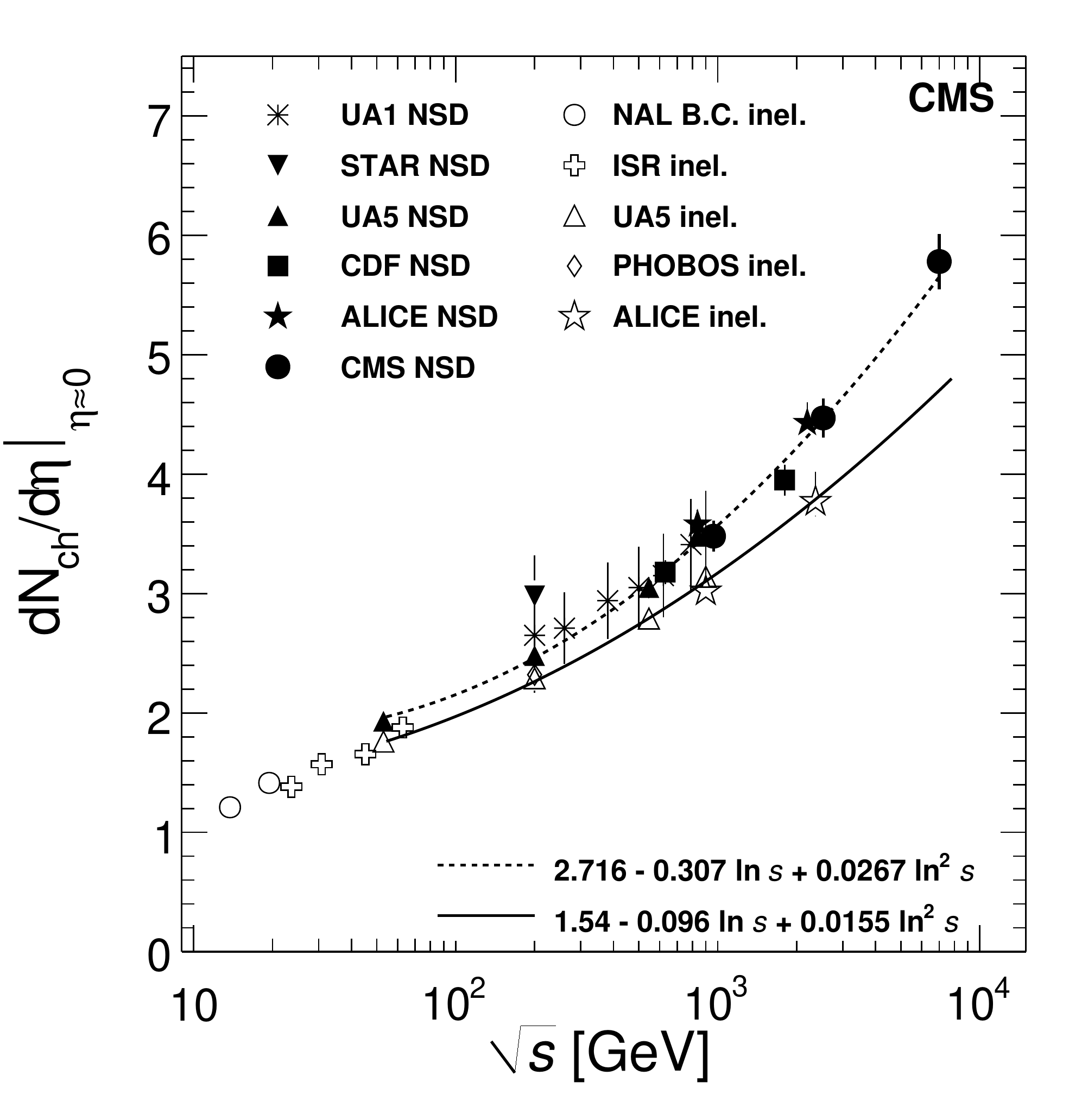}}
\caption{\label{roots_b}
Average value of
\dnchdeta\ in the central $\eta$ region as a function of centre-of-mass
energy in \pp\ and $\pbarp$ collisions.
Also shown are NSD and inelastic measurements from the NAL Bubble
Chamber~\cite{Whitmore:1973ri} (\pbarp), ISR~\cite{Thome:1977ky} (\pp),
UA1~\cite{Albajar:1989an} (\pbarp), UA5~\cite{Alner:1986xu} (\pbarp),
CDF~\cite{Abe:1989td} (\pbarp),
STAR~\cite{star:2008ez} (\pp), PHOBOS~\cite{Nouicer:2004ke} (\pp), and
ALICE~\cite{alice} (\pp). The curves are second-order polynomial
fits for the inelastic (solid) and NSD event selections
(dashed). The error bars include systematic uncertainties, when
available. Data points at 0.9 and 2.36~TeV are slightly displaced
horizontally for visibility.}
\end{figure}

For $|\eta| < 0.5$, the average charged multiplicity density is
$\dnchdeta = 5.78 \pm 0.01\stat\pm 0.23\syst$ for NSD events.
The $\sqrt{s}$ dependence of the measured $\dnchdeta|_{\eta \approx 0}$
is shown in Fig.~\ref{roots_b}, which includes data from various other
experiments. The \dnchdeta\ results reported here show a rather steep
increase between 0.9 and 7~TeV, which is measured to be
($66.1 \pm 1.0 \stat\pm 4.2 \syst$)\%. Using a somewhat different
event selection, the ALICE collaboration has found a similar increase of
($57.6 \pm 0.4 \stat^{+3.6}_{-1.8}\syst$)\%~\cite{alice7tev}. The
measured charged-particle multiplicity is accurate enough to distinguish
among most sets of event-generator tuning parameter values and
various models. The measured value at 7~TeV significantly exceeds the
prediction of 4.57 from PHOJET~\cite{Bopp:1998rc, Engel:1995sb}, and the
predictions of 3.99, 4.18, and 4.34 from the DW~\cite{Bernhard:2010su},
ProQ20~\cite{Buckley:2009bj}, and Perugia0~\cite{Skands:2009zm}
tuning parameter values of PYTHIA, respectively, 
while it is closer to the prediction of 5.48 from the PYTHIA parameter 
set from Ref.~\cite{Moraes:2007rq} and to the recent model predictions 
of 5.58 and 5.78 from Refs.~\cite{Levin:2010dw} and~\cite{Likhoded:2010pc}.
The measured excess of the number of charged hadrons with respect to the
event generators is independent of $\eta$ and concentrated in the 
$\pt<1\GeVc$ range. These differences indicate the need for a 
continued model development and simulation tuning. Work on updated 
event generators based on LHC data is currently underway.

{\it Summary.} Charged-hadron transverse-momentum and pseudorapidity
distributions have been measured in proton-proton collisions at
$\sqrt{s} = 7$~TeV. The numerical values of the data
presented in this paper can be found in the HEPDATA database~\cite{hepdata}.
The combined result for the central pseudorapidity density,
from three mutually consistent methods of measurement, is
\dnchdeta$|_{|\eta| < 0.5} = 5.78\pm 0.01\stat\pm 0.23\syst$
for non-single-diffractive events.
This value is higher than most predictions
and provides new information to constrain ongoing improvements
of soft particle production models and event generators.
The mean transverse momentum has been measured to be
$0.545\pm 0.005\stat\pm 0.015\syst\GeVc$. These studies are the first steps
in the exploration of particle production at the new
centre-of-mass energy frontier, and contribute to the
understanding of the dynamics in soft hadronic interactions.

\noindent {\bf Acknowledgements.}
We congratulate and express our gratitude to our colleagues in the CERN
accelerator departments for the excellent performance of the LHC. We thank
the technical and administrative staff at CERN and other CMS institutes,
and acknowledge support from: FMSR (Austria); FNRS and FWO (Belgium);
CNPq, CAPES, FAPERJ, and FAPESP (Brazil); MES (Bulgaria); CERN; CAS,
MoST, and NSFC (China); COLCIENCIAS (Colombia); MSES (Croatia); RPF
(Cyprus); Academy of Sciences and NICPB (Estonia); Academy of Finland,
ME, and HIP (Finland); CEA and CNRS/IN2P3 (France); BMBF, DFG, and HGF
(Germany); GSRT (Greece); OTKA and NKTH (Hungary); DAE and DST (India);
IPM (Iran); SFI (Ireland); INFN (Italy); NRF and WCU (Korea); LAS (Lithuania);
CINVESTAV, CONACYT, SEP, and UASLP-FAI (Mexico); PAEC (Pakistan); SCSR
(Poland); FCT (Portugal); JINR (Armenia, Belarus, Georgia, Ukraine,
Uzbekistan); MST and MAE (Russia); MSTDS (Serbia); MICINN and CPAN
(Spain); Swiss Funding Agencies (Switzerland); NSC (Taipei); TUBITAK and
TAEK (Turkey); STFC (United Kingdom); DOE and NSF (USA).

\bibliography{auto_generated}

\providecommand{\href}[2]{#2}\begingroup\raggedright\begin{thebibliography}{10}

\bibitem{Evans:2008zzb}
L.~Evans and P.~Bryant~(editors), ``{LHC Machine}'', {\em JINST} {\bf 3} (2008)
  S08001.
\href{http://dx.doi.org/10.1088/1748-0221/3/08/S08001}{{\tt
  doi:10.1088/1748-0221/3/08/S08001}}.

\bibitem{Kittel:2005fu}
W.~Kittel and E.~A. De~Wolf, ``{Soft Multihadron Dynamics}''.
\newblock World Scientific, Singapore, 2005.

\bibitem{1stpaper}
{ CMS} Collaboration, ``{Transverse-momentum and pseudorapidity distributions
  of charged hadrons in $pp$ collisions at $\sqrt{s} = 0.9$ and 2.36 TeV}'',
  {\em JHEP} {\bf 02} (2010) 041.
\href{http://dx.doi.org/10.1007/JHEP02(2010)041}{{\tt
  doi:10.1007/JHEP02(2010)041}}.

\bibitem{JINST}
{ CMS} Collaboration, ``{The CMS experiment at the CERN LHC}'', {\em JINST}
  {\bf 3} (2008) S08004.
  \href{http://dx.doi.org/10.1088/1748-0221/3/08/S08004}{{\tt
  doi:10.1088/1748-0221/3/08/S08004}}.

\bibitem{alice7tev}
{ ALICE} Collaboration, ``{Charged-particle multiplicity measurement in
  proton-proton collisions at $\sqrt{s} = 7$~TeV with ALICE at LHC}'',
\href{http://www.arXiv.org/abs/1004.3514}{{\tt arXiv:1004.3514}}.

\bibitem{GEANT4}
{ Geant4} Collaboration, ``{Geant4: a simulation toolkit}'', {\em Nucl.
  Instrum. and Methods} {\bf A506} (2003) 250.
\href{http://dx.doi.org/10.1016/S0168-9002(03)01368-8}{{\tt
  doi:10.1016/S0168-9002(03)01368-8}}.

\bibitem{Sjostrand:2006za}
T.~Sj\"ostrand, S.~Mrenna, and P.~Skands, ``{PYTHIA 6.4 Physics and Manual;
  v6.420, tune ATLAS}'', {\em JHEP} {\bf 05} (2006)
026.

\bibitem{Moraes:2007rq}
A.~Moraes, C.~Buttar, and I.~Dawson, ``{Prediction for minimum bias and the
  underlying event at LHC energies}'', {\em Eur. Phys. J.} {\bf C50} (2007)
  435.
\href{http://dx.doi.org/10.1140/epjc/s10052-007-0239-1}{{\tt
  doi:10.1140/epjc/s10052-007-0239-1}}.

\bibitem{Bopp:1998rc}
F.~W. Bopp, R.~Engel, and J.~Ranft, ``{Rapidity gaps and the PHOJET Monte
  Carlo, version 1.12-35}'',
\href{http://www.arXiv.org/abs/hep-ph/9803437}{{\tt arXiv:hep-ph/9803437}}.

\bibitem{Engel:1995sb}
R.~Engel, J.~Ranft, and S.~Roesler, ``{Hard diffraction in hadron-hadron
  interactions and in photoproduction}'', {\em Phys. Rev.} {\bf D52} (1995)
  1459.
\href{http://dx.doi.org/10.1103/PhysRevD.52.1459}{{\tt
  doi:10.1103/PhysRevD.52.1459}}.

\bibitem{Sikler:2009nx}
F.~Sikl\'er, ``{Study of clustering methods to improve primary vertex finding
  for collider detectors}'', \href{http://www.arXiv.org/abs/0911.2767}{{\tt
  arXiv:0911.2767}}.
\href{http://dx.doi.org/10.1016/j.nima.2010.04.058}{{\tt
  doi:10.1016/j.nima.2010.04.058}}.

\bibitem{Tsallis:1987eu}
C.~Tsallis, ``{Possible generalization of Boltzmann-Gibbs statistics}'', {\em
  J. Stat. Phys.} {\bf 52} (1988) 479.
  \href{http://dx.doi.org/10.1007/BF01016429}{{\tt doi:10.1007/BF01016429}}.

\bibitem{Wilk:2008ue}
G.~Wilk and Z.~W{\l}odarczyk, ``{Power laws in elementary and heavy-ion
  collisions: A Story of fluctuations and nonextensivity?}'', {\em Eur. Phys.
  J.} {\bf A40} (2009) 299.
\href{http://dx.doi.org/10.1140/epja/i2009-10803-9}{{\tt
  doi:10.1140/epja/i2009-10803-9}}.

\bibitem{Biro:2008hz}
T.~S. Bir\'o, G.~Purcsel, and K.~{\"U}rm{\"o}ssy, ``{Non-Extensive Approach to
  Quark Matter}'', {\em Eur. Phys. J.} {\bf A40} (2009) 325,
  \href{http://www.arXiv.org/abs/0812.2104}{{\tt arXiv:0812.2104}}.
\href{http://dx.doi.org/10.1140/epja/i2009-10806-6}{{\tt
  doi:10.1140/epja/i2009-10806-6}}.

\bibitem{Alner:1986xu}
{ UA5} Collaboration, ``{Scaling of pseudorapidity distributions at c.m.
  energies up to 0.9 TeV}'', {\em Z. Phys.} {\bf C33} (1986) 1.
  \href{http://dx.doi.org/10.1007/BF01410446}{{\tt doi:10.1007/BF01410446}}.

\bibitem{alice}
{ ALICE} Collaboration, ``{Charged-particle multiplicity measurement in
  proton-proton collisions at $\sqrt{s} = 0.9$ and 2.36 TeV with ALICE at
  LHC}'',
\href{http://www.arXiv.org/abs/1004.3034}{{\tt arXiv:1004.3034}}.

\bibitem{Aad:2010rd}
{ ATLAS} Collaboration, ``{Charged-particle multiplicities in $pp$ interactions
  at $\sqrt{s} = 900$~GeV measured with the ATLAS detector at the LHC}'', {\em
  Phys. Lett.} {\bf B688} (2010) 21,
  \href{http://www.arXiv.org/abs/1003.3124}{{\tt arXiv:1003.3124}}.
\href{http://dx.doi.org/10.1016/j.physletb.2010.03.064}{{\tt
  doi:10.1016/j.physletb.2010.03.064}}.

\bibitem{Albajar:1989an}
{ UA1} Collaboration, ``{A Study of the General Characteristics of
  Proton-Antiproton Collisions at 0.2 to 0.9 TeV}'', {\em Nucl. Phys.} {\bf
  B335} (1990) 261.
\href{http://dx.doi.org/10.1016/0550-3213(90)90493-W}{{\tt
  doi:10.1016/0550-3213(90)90493-W}}.

\bibitem{Rossi:1974if}
A.~M. Rossi {et~al.}, ``{Experimental Study of the Energy Dependence in Proton
  Proton Inclusive Reactions}'', {\em Nucl. Phys.} {\bf B84} (1975) 269.
\href{http://dx.doi.org/10.1016/0550-3213(75)90307-7}{{\tt
  doi:10.1016/0550-3213(75)90307-7}}.

\bibitem{Alexopoulos:1988na}
{ E735} Collaboration, ``{Multiplicity dependence of the transverse momentum
  spectrum for centrally produced hadrons in antiproton-proton collisions at
  $\sqrt{s}$ = 1.8 TeV}'', {\em Phys. Rev. Lett.} {\bf 60} (1988) 1622.
\href{http://dx.doi.org/10.1103/PhysRevLett.60.1622}{{\tt
  doi:10.1103/PhysRevLett.60.1622}}.

\bibitem{Abe:1988yu}
{ CDF} Collaboration, ``{Transverse momentum distributions of charged particles
  produced in \pbarp\ interactions at $\sqrt{s} = 630$ GeV and 1800 GeV}'',
  {\em Phys. Rev. Lett.} {\bf 61} (1988) 1819.
\href{http://dx.doi.org/10.1103/PhysRevLett.61.1819}{{\tt
  doi:10.1103/PhysRevLett.61.1819}}.

\bibitem{Bernhard:2010su}
R.~Bernhard {et~al.}, ``{Proceedings of the First International Workshop on
  Multiple Partonic Interactions at the LHC (MPI08)}'',
\href{http://www.arXiv.org/abs/1003.4220}{{\tt arXiv:1003.4220}}.

\bibitem{Buckley:2009bj}
A.~Buckley {et~al.}, ``{Systematic event generator tuning for the LHC}'', {\em
  Eur. Phys. J.} {\bf C65} (2010) 331,
  \href{http://www.arXiv.org/abs/0907.2973}{{\tt arXiv:0907.2973}}.
\href{http://dx.doi.org/10.1140/epjc/s10052-009-1196-7}{{\tt
  doi:10.1140/epjc/s10052-009-1196-7}}.

\bibitem{Skands:2009zm}
P.~Z. Skands, ``{The Perugia Tunes}'',
\href{http://www.arXiv.org/abs/0905.3418}{{\tt arXiv:0905.3418}}.

\bibitem{Levin:2010dw}
E.~Levin and A.~H. Rezaeian, ``{Gluon saturation and inclusive hadron
  production at LHC}'',
\href{http://www.arXiv.org/abs/1005.0631}{{\tt arXiv:1005.0631}}.

\bibitem{Likhoded:2010pc}
A.~K. Likhoded, A.~V. Luchinsky, and A.~A. Novoselov, ``{Light hadron
  production in semi-inclusive $pp$-scattering at LHC}'',
  \href{http://www.arXiv.org/abs/1005.1827}{{\tt arXiv:1005.1827}}.

\bibitem{Whitmore:1973ri}
J.~Whitmore, ``{Experimental Results on Strong Interactions in the NAL Hydrogen
  Bubble Chamber}'', {\em Phys. Rept.} {\bf 10} (1974) 273.
\href{http://dx.doi.org/10.1016/0370-1573(74)90046-5}{{\tt
  doi:10.1016/0370-1573(74)90046-5}}.

\bibitem{Thome:1977ky}
{ Aachen-CERN-Heidelberg-Munich} Collaboration, ``{Charged Particle
  Multiplicity Distributions in $pp$ Collisions at ISR Energies}'', {\em Nucl.
  Phys.} {\bf B129} (1977) 365.
\href{http://dx.doi.org/10.1016/0550-3213(77)90122-5}{{\tt
  doi:10.1016/0550-3213(77)90122-5}}.

\bibitem{Abe:1989td}
{ CDF} Collaboration, ``{Pseudorapidity distributions of charged particles
  produced in \pbarp\ interactions at $\sqrt{s} = 630$ GeV and 1800 GeV}'',
  {\em Phys. Rev.} {\bf D41} (1990) 2330.
\href{http://dx.doi.org/10.1103/PhysRevD.41.2330}{{\tt
  doi:10.1103/PhysRevD.41.2330}}.

\bibitem{star:2008ez}
{ STAR} Collaboration, ``{Systematic measurements of identified particle
  spectra in $pp$, d-Au and Au-Au collisions at the STAR detector}'', {\em
  Phys. Rev.} {\bf C79} (2009) 034909.
\href{http://dx.doi.org/10.1103/PhysRevC.79.034909}{{\tt
  doi:10.1103/PhysRevC.79.034909}}.

\bibitem{Nouicer:2004ke}
{ PHOBOS} Collaboration, ``{Pseudorapidity distributions of charged particles
  in $dAu$ and $pp$ collisions at $\sqrt{s_{\rm NN}}=200$ GeV}'', {\em J.
  Phys.} {\bf G30} (2004) S1133.
\href{http://dx.doi.org/10.1088/0954-3899/30/8/075}{{\tt
  doi:10.1088/0954-3899/30/8/075}}.

\bibitem{hepdata}
{ {CMS}} Collaboration. Available at HEPDATA:
  \url{http://hepdata.cedar.ac.uk/view/irn8656010}, 2010.

\end{thebibliography}\endgroup
\cleardoublepage\appendix\section{The CMS Collaboration \label{app:collab}}\begin{sloppypar}\hyphenpenalty=5000\widowpenalty=500\clubpenalty=5000\textbf{Yerevan Physics Institute,  Yerevan,  Armenia}\\*[0pt]
V.~Khachatryan, A.M.~Sirunyan, A.~Tumasyan
\vskip\cmsinstskip
\textbf{Institut f\"{u}r Hochenergiephysik der OeAW,  Wien,  Austria}\\*[0pt]
W.~Adam, T.~Bergauer, M.~Dragicevic, J.~Er\"{o}, C.~Fabjan, M.~Friedl, R.~Fr\"{u}hwirth, V.M.~Ghete, J.~Hammer\cmsAuthorMark{1}, S.~H\"{a}nsel, M.~Hoch, N.~H\"{o}rmann, J.~Hrubec, M.~Jeitler, G.~Kasieczka, W.~Kiesenhofer, M.~Krammer, D.~Liko, I.~Mikulec, M.~Pernicka, H.~Rohringer, R.~Sch\"{o}fbeck, J.~Strauss, A.~Taurok, F.~Teischinger, W.~Waltenberger, G.~Walzel, E.~Widl, C.-E.~Wulz
% ...
\vskip\cmsinstskip
% ...
1:~~Also at CERN, European Organization for Nuclear Research, Geneva, Switzerland\\
% ...

\end{sloppypar}
\end{document}